\documentclass[10pt]{article}
\usepackage{latexsym}
\usepackage{amssymb}
\usepackage{amsmath}
\usepackage{amscd}
\usepackage{amsthm}
\usepackage[left=2cm,top=2.5cm,right=2.5cm,bottom=1.5cm]{geometry}

\usepackage[dvips]{graphicx}
\usepackage{hyperref}

\begin{document}
\begin{center}
\Large{\bf{Some Anisotropic Dark Energy Models in Bianchi Type-V Space-time}}
\\
\vspace{10mm} \normalsize{Anil Kumar Yadav}\\ \vspace{4mm} 
\normalsize{Department of Physics, Anand Engineering
College, Keetham, Agra-282 007, India} \\
\vspace{2mm}
\normalsize{E-mail: abanilyadav@yahoo.co.in}\\
\end{center}
\begin{abstract}
The paper deals  with Bianchi type V Universe, which has dynamical energy 
density. We consider Bianchi type V space-time, introducing three different skewness parameters along 
spatial directions to quantify the deviation of pressure from isotropy. 
To study the anisotropic nature of the dynamical dark energy, we assume that 
the skewness parameters are time dependent. It is found that the Universe achieves flatness
in quintessence model. The physical behavior of the
Universe has been discussed in detail.
\end{abstract}


\smallskip
\vspace{4mm}
\section{Introduction}
Recent observations have revolutionized our understanding of cosmology. 
Analysis of type Ia supernovae (SN Ia) (Perlmutter et al. 1997, 1998, 1999; Riess et al. 1998, 2004), 
cosmic microwave background (CMB) anisotropy (Caldwell et al 2002; Huang et al. 2006),
and large scale structure (Daniel et al. 2008) strongly indicate that dark energy (DE) dominates the present Universe, 
causing cosmic acceleration. This acceleration is realized with negative pressure and positive energy 
density that violate the strong energy condition. This violation gives a reverse gravitational effect. 
Due to this effect, the Universe gets a jerk and the transition from the earlier deceleration phase to the recent 
acceleration phase  take place (Caldwell et al. 2006). The cause of this sudden transition and the source of accelerated 
expansion is still unknown. In physical cosmology and astronomy, the simplest candidate
for the DE is the cosmological constant ($\Lambda$), but it needs to be extremely fine-tuned 
to satisfy the current value of the DE density, which is a serious problem. Alternatively, to explain 
the decay of the density, the
different forms of dynamically changing DE with an effective
equation of state (EoS), $\omega=p^{(de)}/\rho^{(de)}<-1/3$, were
proposed instead of the constant vacuum energy density. Other
possible forms of DE include quintessence ($\omega>-1$) (Steinhardt et al. 1999),
phantom ($\omega<-1$) (Caldwell et al. 2002) etc. While the the possibility
$\omega<<-1$ is ruled out by current cosmological data from SN Ia
(Supernovae Legacy Survey, Gold sample of Hubble Space Telescope)
(Riess et al. 2004; Astier et al. 2006), CMBR (WMAP, BOOMERANG) (Eisentein et al. 2005; MacTavish et al. 2006) 
and large scale
structure (Sloan Digital Sky Survey) (Komatsu et al. 2009) data, the dynamically
evolving DE crossing the phantom divide line
(PDL) ($\omega=-1$) is mildly favored.\\
\indent The anisotropy of the DE within the framework of Bianchi
type space-times is found to be useful in generating arbitrary
ellipsoidality to the Universe, and to fine tune the observed CMBR
anisotropies. Koivisto and Mota (2008a,2008b) have investigated cosmological model 
with anisotropic EoS. They have proposed a different approach to resolve CMB anisotropy 
problem; even if the CMB formed isotropically at early time, it could be distorted by the direction 
dependent acceleration of the future Universe in such a way that it appears to us anomalous 
at the largest scales. They have investigated a cosmological model containing a DE component 
which has a non dynamical anisotropic EoS and interacts with the perfect fluid component. 
They have also suggested that cosmological models with anisotropic EoS can explain the quadrupole 
problem and can be tested by SN Ia data. Kumar and Singh (2010) have studied Bianchi type I cosmological 
models with constant deceleration parameter (DP) in the presence of anisotropic DE and perfect fluid. 
They have considered phenomenological parametrization of
minimally interacting DE in terms of its EoS and time-dependent
skewness parameters $(\delta(t),\gamma(t), \eta(t))$. 
Leon and Sarikadis (2010) have investigated that anisotropic geometries in modified 
gravitational frameworks present radically difference cosmological behaviors comparing to the simple isotropic 
scenarios. Akarsu et al. (2010) have investigated 
Bianchi-I anisotropic DE model with constant DP. 
Yadav and Yadav (2011a), Yadav et al. (2011b) 
have studied anisotropic DE models with variable EoS parameter. They have suggested that the dynamics of 
EoS parameter describe the present acceleration of Universe
 i.e. from earlier deceleration phase to recent acceleration phase. Recently Pradhan et al (2011a, 2011b)
have studied anisotropic DE models in different physical contexts. They have found that in the earlier stage 
EoS parameter was positive and it evolves with negative sign at present epoch.   
  
Bianchi type-V Universe is generalization of the open Universe in FRW cosmology and 
hence it's study is important in
the study of DE models in  Universe with non-zero curvature (1994). A number of authors such as
Collins (1974), Maartens and Nel (1978), Wrainwright et al (1979), Canci et al (2001),
Pradhan et al (2004), Singh et al. (2008), Yadav (2009) have studied Bianchi type-V model 
in different physical contexts.
Recently Kumar and Yadav (2010) have studied isotropic DE model with variable EoS parameter in 
Bianchi type V space-time and found that the Universe is dominated by DE at 
present epoch and after dominance of DE, Universe achieves flatness. Following Eriksen et al (2004),
it is found that some large-angle anomalies appear in CMB radiations which violate the statistical isotropy 
of the Universe. This motivates the researcher to consider the model of Universe with anisotropic DE.

In this paper, we have studied some physically realistic and totally anisotropic 
Bianchi-V models with anisotropic DE and perfect fluid. To study the anisotropic nature of 
DE, we have assumed the time dependent skewness parameter, which modify EoS. 
The time dependent forms of the skewness parameter provide exact solutions of Einstein's field equation 
together with the special law of variation of Hubble's parameter. The paper is
organized as follows. In Section 2, the models and field equations
have been presented. The Section 3 deals with the exact solutions of
the field equations and physical behavior of the models. Finally,
the results are discussed in section 4.

\section{Model and field equations}
The spatially homogeneous and anisotropic Bianca-V space-time is
described by the line element
\begin{equation}\label{eq1}
ds^{2} =-dt^{2} +A^{2}dx^{2} +e^{2\alpha x}(B^{2}dy^{2}
+C^{2}dz^{2}),
\end{equation}
where  $A$ ,  $B$ and  $C$ are the metric functions of cosmic time
$t$ and $\alpha$ is a constant.

We define $a=(ABC)^{\frac{1}{3} } $ as the average scale factor of
the space-time (\ref{eq1}) so that the average Hubble's parameter
reads as
\begin{equation}\label{eq2}
H=\frac{\dot{a}}{a} =\frac{1}{3} \left(\frac{\dot{A}}{A}
+\frac{\dot{B}}{B} +\frac{\dot{C}}{C} \right),
\end{equation}
where $a=(ABC)^{\frac{1}{3} } $  is the average scale factor and an
over dot denotes derivative with respect to the cosmic time  $t$ .\\
\indent The directional Hubble parameters along $x$ , $y$ and $z$
coordinate axes, respectively, may be defined as
\begin{equation}\label{eq3}
 H_{x} =\frac{\dot{A}}{A},\;\;\;\;  H_{y} =\frac{\dot{B}}{B},\;\;\;\;H_{z}
 =\frac{\dot{C}}{C}.
\end{equation}
The Einstein's field equations (in gravitational units $8\pi G=c=1$) read as
\begin{equation}\label{eq4}
R^{\;i}_{\;j}-\frac{1}{2} g^{\;i}_{\;j}R = - T^{(m)\;i}_{\;\;j}- T^{(de)\;i}_{\;\;j},
\end{equation}
 where $T^{(m)\;i}_{\;\;j}$ and $T^{(de)\;i}_{\;j}$ are the energy momentum
tensors of perfect fluid and DE, respectively. These are given by
\begin{equation}\label{eq5}
T^{(m)\;i}_{\;\;j}
=\text{diag}\;[-\rho^{(m)},\;p^{(m)}\;,p^{(m)},\;p^{(m)}]
\end{equation}
and
\begin{eqnarray}\label{eq6}
T^{(de)\;i}_{\;\;j}&=&\text{diag}\;[-\rho^{(de)},\;p_{x}^{(de)},\;p_{y}^{(de)},\;p_{z}^{(de)}]\nonumber\\
                   &=& \text{diag}\;[-1,\;\omega_{x},\;\omega_{y},\;\omega_{z}]\rho^{(de)}\nonumber\\
                   &=&\text{diag}\;[-1,\;w+\delta,\;w+\gamma,\;w+\eta]\rho^{(de)}
\end{eqnarray}
where $\rho^{(m)}$ and $p^{(m)}$ are, respectively the energy
density and pressure of the perfect fluid component; $\rho^{(de)}$
is the energy density of the DE component; $\delta(t)$, $\gamma(t)$
and $\eta(t)$ are skewness parameters, which modify EoS (hence
pressure) of the DE component and are functions of the cosmic time
$t$; $\omega$ is the EoS parameter of DE; $\omega_{x}$, $\omega_{y}$
and $\omega_{z}$ are the directional EoS parameters along $x$ , $y$
and $z$ coordinate axes, respectively and we assume the four
velocity vector
$u^{i}=(1,0,0,0)$ satisfying $ u^{i} u_{i} =-1$.\\
\indent In a co moving coordinate system ($u^{i}=\delta_{0}^{i}$),
the field equations (\ref{eq4}), for the anisotropic Bianchi type-I
space-time (\ref{eq1}),
in case of (\ref{eq5}) and (\ref{eq6}), read as\\
\begin{equation}\label{eq7}
\frac{\ddot{B}}{B} +\frac{\ddot{C}}{C}+\frac{\dot{B}\dot{C}}{BC}
-\frac{\alpha^{2}}{A^{2}}=-p^{(m)}-(\omega+\delta)\rho^{(de)},
\end{equation}
\begin{equation}\label{eq8}
\frac{\ddot{C}}{C} +\frac{\ddot{A}}{A}
+\frac{\dot{C}\dot{A}}{CA}-\frac{\alpha^{2}}{A^{2}}
=-p^{(m)}-(\omega+\gamma)\rho^{(de)},
\end{equation}
\begin{equation}\label{eq9}
\frac{\ddot{A}}{A} +\frac{\ddot{B}}{B}
+\frac{\dot{A}\dot{B}}{AB}-\frac{\alpha^{2}}{A^{2}}
=-p^{(m)}-(\omega+\eta)\rho^{(de)},
\end{equation}
\begin{equation}\label{eq10}
\frac{\dot{A}\dot{B}}{AB} +\frac{\dot{B}\dot{C}}{BC}
+\frac{\dot{C}\dot{A}}{CA}-\frac{3\alpha^{2}}{A^{2}}
 =\rho^{(m)}+\rho^{(de)}.
\end{equation}
\begin{equation}\label{eq11}
2\frac{\dot{A}}{A}-\frac{\dot{B}}{B}
-\frac{\dot{C}}{C}=0
\end{equation}
\indent We assume that the perfect fluid and DE components interact
minimally.  Therefore, the energy momentum tensors of the
two sources may be conserved separately.\\
\indent The energy conservation equation
$T^{(m)\;j}_{\;\;\;\;\;\;\;\;\;\;;j} =0$, of the perfect fluid
leads to
\begin{equation}\label{eq12}
\dot\rho^{(m)}+3(\rho^{(m)}+p^{(m)})H=0,
\end{equation}
whereas the energy conservation equation
$T^{(de)\;j}_{\;\;\;\;\;\;\;\;\;\;;j} =0$, of the DE component
yields
\begin{equation}\label{eq13}
\dot\rho^{(de)}+3\rho^{(de)}(\omega+1)H +\rho^{(de)}\left(\delta
H_{x}+\gamma H_{y}+\eta H_{z}\right)=0,
\end{equation}
where we have used the equation of state
 $p^{(de)}=\omega\rho^{(de)}$.\\

Equations (\ref{eq7})-(\ref{eq10}) can be written in terms of $H$, $\sigma$ and $q$ as
\begin{equation}
\label{eq14}
p^{(de)}+\frac{1}{3}\left(3\omega + \delta +\gamma +\eta\right)\rho^{(de)}=H^{2}(2q - 1) - \sigma^{2} +\frac{\alpha^{2}}{A^{2}}
\end{equation}
\begin{equation}
\label{eq15}
\rho^{(m)}+\rho^{(de)}=3H^{2}-\sigma^{2} -\frac{3\alpha^{2}}{A^{2}}
\end{equation}
Where $q$ and $\sigma$ are deceleration parameter and shear scalar respectively.

\section{Solution of Field Equations}
\indent Berman (1983), Berman and Gomide (1988), recently Kumar (2010) obtained
some FRW cosmological models with constant DP
and showed that the constant DP models stand
adequately for our present view of different phases of the evolution
of Universe. In this paper, we show how the constant deceleration
parameter models with metric (\ref{eq1}) behave in the presence of 
anisotropic DE. According to the law,
the variation of the average Hubble parameter is given by Singh et al (2008)
\begin{equation}\label{eq16}
H=Da^{-n},
\end{equation}
where  $D>0$  and  $n\geq 0$ are constants.\\

Following, Akarsu and Kilinc (2010), we split the conservation of energy momentum tensor 
of the DE into two parts, One corresponds to deviations of EoS parameter and other is the 
deviation-free part of $T^{(de)\;ij}_{\;\;\;\;\;\;\;\;\;\;;j} =0$:
\begin{equation}
\label{eq17}
\dot{\rho}+3\rho^{(de)}\left(\omega+1\right)H = 0
\end{equation}
and
\begin{equation}
\label{eq18}
\rho^{(de)}\left(\delta
H_{x}+\gamma H_{y}+\eta H_{z}\right)=0,
\end{equation}
According to equations (\ref{eq17}) and (\ref{eq18}) the behaviour of 
$\rho^{(de)}$ is controlled by the deviation-free part of EoS parameter of DE but 
deviations will affect $\rho^{(de)}$ indirectly, since, as can be seen later, they affect the 
value of EoS parameter. Of course, the choice of skewness parameters are quite arbitrary but, since we are 
looking for a physically viable models of Universe consistent with observations. We consider 
the skewness parameters $\delta$, $\gamma$ and $\eta$ be function of cosmic time and we constrained 
$\delta$, $\gamma$ and $\eta$ by assuming a special dynamics which is consistent with eq. (\ref{eq18}). 
The dynamics of skewness parameters on x-axis, y-axis and z-axis are assumed to be\\
\begin{equation}\label{eq19}
\delta(t)=\alpha\left(H_{y}+H_{z}\right)\frac{1}{\rho^{(de)}},
\end{equation}
\begin{equation}\label{eq20}
\gamma(t)=-\alpha H_{x}\frac{1}{\rho^{(de)}},
\end{equation}
\begin{equation}\label{eq21}
\eta(t)=-\alpha H_{x}\frac{1}{\rho^{(de)}},
\end{equation}
where $\alpha$ is an arbitrary constant, which parameterizes the
anisotropy of the DE. In literature, many authors have considered
totally anisotropic Bianchi-I, Bianchi-III and Bianchi-V space-times with only two skewness
parameters of DE (Akarsu et al. 2010; Yadav et al. 2011a, 2011b; Pradhan et al. 2011a, 2011b).

Finally, we assume that $\omega=$ const. so that we can study
different models related to the DE by choosing different values of
$\omega$, viz. phantom ($\omega<-1$), cosmological constant
($\omega=-1$) and quintessence
($\omega>-1$).\\
\indent In view of the assumptions (\ref{eq19})-(\ref{eq21}) and
$\omega=$ const., equation (\ref{eq11}) can be integrated to obtain
\begin{equation}\label{eq22}
\rho^{(de)}\;(t)=\rho_{0}a^{-3(\omega+1)},
\end{equation}
where $\rho_{0}$ is a positive constant of integration.\\
Integrating (\ref{eq11}) and absorbing the constant of integration in
$B$ or $C$, without loss of generality, we obtain
\begin{equation}\label{eq23}
A^{2} =BC.
\end{equation}
Subtracting (\ref{eq7}) from (\ref{eq8}), (\ref{eq7}) from
(\ref{eq9}), (\ref{eq8}) from (\ref{eq10}) and taking second integral
of each, we get the following three relations respectively:
\begin{equation}\label{eq24}
\frac{A}{B} =d_{1} \exp \left[ x_{1}\int a^{-3} dt
-\frac{\alpha\rho_{0}}{\omega}\int a^{-3(\omega+1)}dt\right],
\end{equation}
\begin{equation}\label{eq25}
\frac{A}{C} =d_{2} \exp \left[ x_{2}\int a^{-3} dt
-\frac{\alpha\rho_{0}}{\omega}\int a^{-3(\omega+1)}dt\right],
\end{equation}
\begin{equation}\label{eq26}
\frac{B}{C} =d_{3} \exp \left(x_{3} \int a^{-3} dt \right),
\end{equation}
where  $d_{1} $ , $x_{1} $ ,  $d_{2} $ ,  $x_{2} $ ,  $d_{3} $  and
$x_{3} $ are constants of integration.
From equations (\ref{eq24})-(\ref{eq26}) and (\ref{eq23}), the metric functions can
be explicitly written as
\begin{equation}\label{eq27}
A(t) =a \exp \left[
-\frac{2\alpha\rho_{0}}{3\omega}\int a^{-3(\omega+1)}dt\right],
\end{equation}
\begin{equation}\label{eq28}
B(t) =m a \exp \left[ l\int a^{-3} dt
+\frac{\alpha\rho_{0}}{3\omega}\int a^{-3(\omega+1)}dt\right],
\end{equation}
\begin{equation}\label{eq29}
C(t) =m^{-1} a \exp \left[ -l\int a^{-3} dt
+\frac{\alpha\rho^{(de)}_{0}}{3\omega}\int a^{-3(\omega+1)}dt\right],
\end{equation}
where
\begin{equation}
\label{eq30}
m =\sqrt[{3}]{d_{2} d_{3} } ,~~  l =\frac{(x_{2} +x_{3} )}{3}
\end{equation}
with
\begin{equation}\label{eq31}
d_{2}=d_{1}^{-1} ,~~  x_{2} =-x_{1}.
\end{equation}

\indent In the following subsections, we discuss the cosmologies for $n\neq 0$ and $n=0$, respectively.\\
\subsection{DE Cosmology for $n\neq 0$}
In this case, integration of (\ref{eq16}) leads to
\begin{equation}\label{eq32}
a(t)=(nDt)^{\frac{1}{n}},
\end{equation}
where the constant of integration has been omitted by assuming that
$a=0 $ at $t=0 $.\\
\noindent Using (\ref{eq32}) into (\ref{eq27})-(\ref{eq29}), we get the
following expressions for scale factors:
\begin{equation}\label{eq33}
A(t)= (nDt)^{\frac{1}{n} } \exp \left[-\frac{2\alpha\rho_{0}}{3\omega
D(n-3\omega-3)}(nDt)^{\frac{n-3\omega-3}{n}} \right] ,
\end{equation}
\begin{equation}\label{eq34}
B(t)=m (nDt)^{\frac{1}{n} } \exp \left[\frac{l }{D(n-3)}
(nDt)^{\frac{n-3}{n} } +\frac{\alpha\rho_{0}}{3\omega
D(n-3\omega-3)}(nDt)^{\frac{n-3\omega-3}{n}}\right],
\end{equation}
\begin{equation}\label{eq35}
C(t)=m^{-1} (nDt)^{\frac{1}{n} } \exp \left[\frac{-l }{D(n-3)}
(nDt)^{\frac{n-3}{n} }+\frac{\alpha\rho_{0}}{3\omega
D(n-3\omega-3)}(nDt)^{\frac{n-3\omega-3}{n}}\right],
\end{equation}
where $n\neq3$.\\
\indent The physical parameters such as directional Hubble
parameters $(H_{x},H_{y},H_{z})$, average Hubble parameter $(H)$,
 anisotropy parameter ($\bar{A}$), expansion scalar ($\theta$) and spatial volume ($V$)
are, respectively, given by
\begin{equation}\label{eq36}
H_{x} =(nt)^{-1}-\frac{2\alpha\rho_{0}}{3\omega}(nDt)^{\frac{-3(\omega+1)}{n}},
\end{equation}
\begin{equation}
\label{eq37}
H_{y} =(nt)^{-1} + l (nDt )^{\frac{-3}{n}
}+\frac{\alpha\rho_{0}}{3\omega}(nDt)^{\frac{-3(\omega+1)}{n}},
\end{equation}
\begin{equation}
\label{eq38}
H_{z} =(nt)^{-1} -l (nDt )^{\frac{-3}{n}
}+\frac{\alpha\rho_{0}}{3\omega}(nDt)^{\frac{-3(\omega+1)}{n}},
\end{equation}
\begin{equation}
\label{eq39}
H=(nt)^{-1},
\end{equation}
\begin{eqnarray}\label{eq40}
\bar{A} &=&\frac{1}{3}\left[\left(\frac{H_{x} -H}{H} \right)
^{2}+\left(\frac{H_{y} -H}{H} \right) ^{2}+\left(\frac{H_{z} -H}{H}
\right) ^{2}\right]  \nonumber\\ &=&\frac{2}{3 D^2}
\left[l^{2}(nDt)^{\frac{2(n-3)}{n}
} +\frac{\alpha^{2}{\rho_{0}}^{2}}{3\omega^{2}}(nDt)^{\frac{2(n-3\omega-3)}{n}}\right],
\end{eqnarray}
\begin{equation}\label{eq41}
\theta =u^{i}_{;i} =\frac{3\dot{a}}{a}=3(nt)^{-1},
\end{equation}
\begin{equation}\label{eq42}
V =(nDt )^{\frac{3}{n}} \exp{(2\alpha x)},
\end{equation}

Shear scalar of the model reads as
\begin{equation}\label{43}
\sigma^{2} =l^{2}(nDt)^{\frac{-6}{n}
}+\frac{\alpha^{2}{\rho_{0}}^{2}}{3\omega^{2}}(nDt)^{\frac{-6(\omega+1)}{n}}.
\end{equation}

The value of DP ($q$) is found to be
\begin{equation}\label{eq44}
q=-\frac{a\ddot{a}}{{\dot{a}}^{2}} = n-1,
\end{equation}
which is a constant. The sign of $q$ indicates whether the model
inflates or not. A positive sign of $q$, i.e., $n>1$ corresponds to
the standard decelerating model whereas the negative sign of $q$,
i.e., $0< n<1$ indicates acceleration. The expansion of the Universe at
a constant rate corresponds to $n=1$, i.e., $q=0$. Also, recent
observations of SN Ia, reveal that the
present Universe is accelerating and value of DP lies somewhere in
the range $-1<q< 0.$ It follows that in the derived model, one can
choose the values of DP consistent with the observations.\\
\indent The skewness parameters of DE are as follows:
\begin{equation}\label{45}
\delta(t)=\frac{\alpha}{\rho_{0}}\left[2D(Dnt)^{-\frac{3(\omega+1)-n}{n}}+\frac{2\alpha\rho_{0}}{3\omega}
\right],
\end{equation}
\begin{equation}\label{46}
\gamma(t)=\eta(t)=-\frac{\alpha}{\rho_{0}}\left[2D(Dnt)^{-\frac{3(\omega+1)-n}{n}}-\frac{2\alpha\rho_{0}}{3\omega}
\right].
\end{equation}
\indent In view of (\ref{eq5}), the directional EoS parameters of DE
are given by
\begin{equation}\label{47}
\omega_{x}=\omega+\frac{\alpha}{\rho_{0}}\left[2D(Dnt)^{-\frac{3(\omega+1)-n}{n}}+\frac{2\alpha\rho_{0}}{3\omega}
\right],
\end{equation}
\begin{equation}\label{48}
\omega_{y}=\omega_{z}=\omega-\frac{\alpha}{\rho_{0}}\left[2D(Dnt)^{-\frac{3(\omega+1)-n}{n}}-\frac{2\alpha\rho_{0}}{3\omega}\right].
\end{equation}
\indent The energy density and pressure of the DE components are
obtained as
\begin{equation}\label{49}
\rho^{(de)}=\rho_{0}(nDt)^{\frac{-3(\omega+1)}{n}},
\end{equation}
\begin{equation}\label{50}
p^{(de)}=\omega\rho_{0}(nDt)^{\frac{-3(\omega+1)}{n}}.
\end{equation}
\begin{figure}
\begin{center}
\includegraphics [height=6 cm]{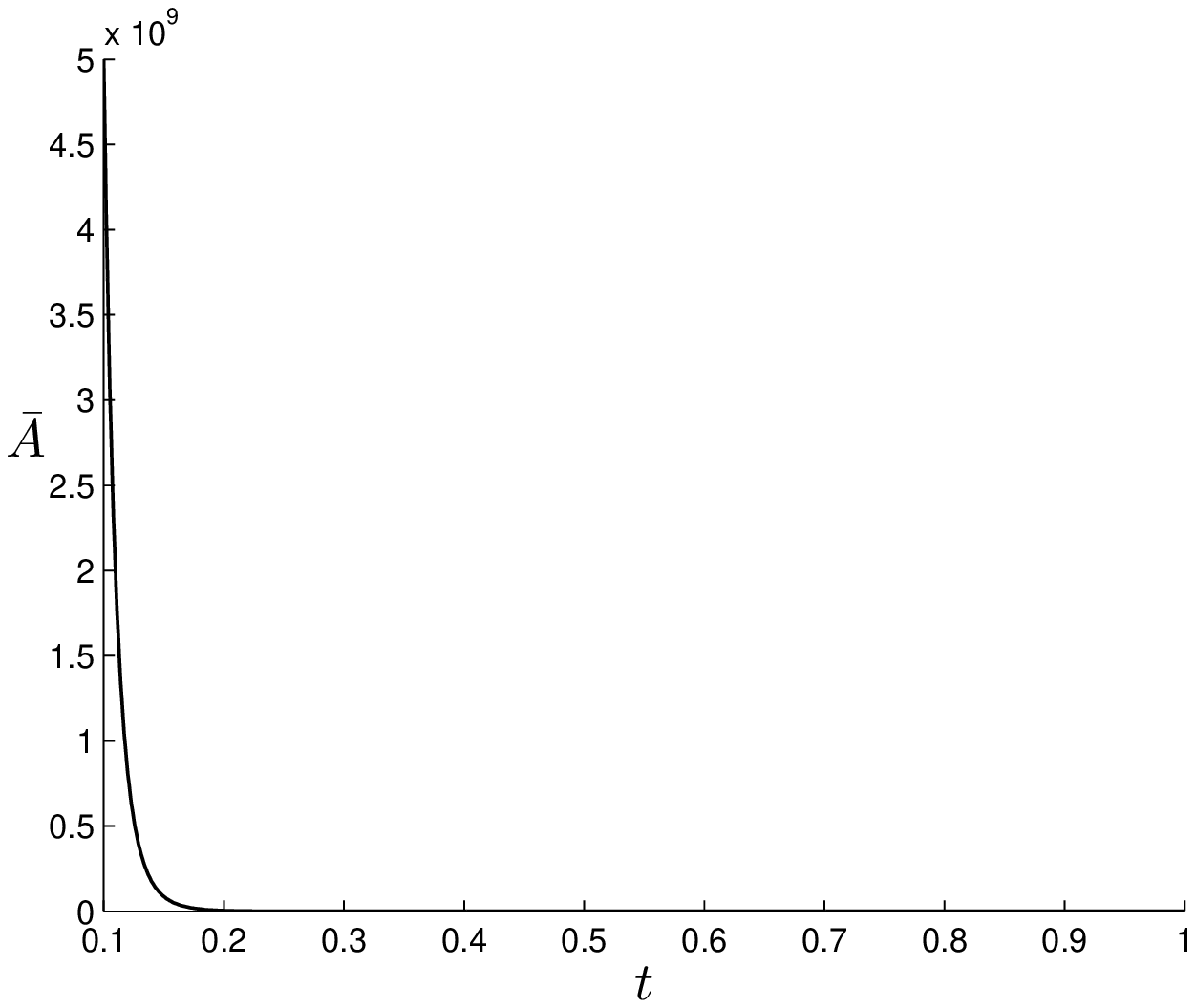}
\caption{Plot of anisotropic parameter $(\bar{A})$ versus time (t).} \label{fg:anil32F1.eps}
\end{center}
\end{figure}
\begin{figure}
\begin{center}
\includegraphics [height=6 cm]{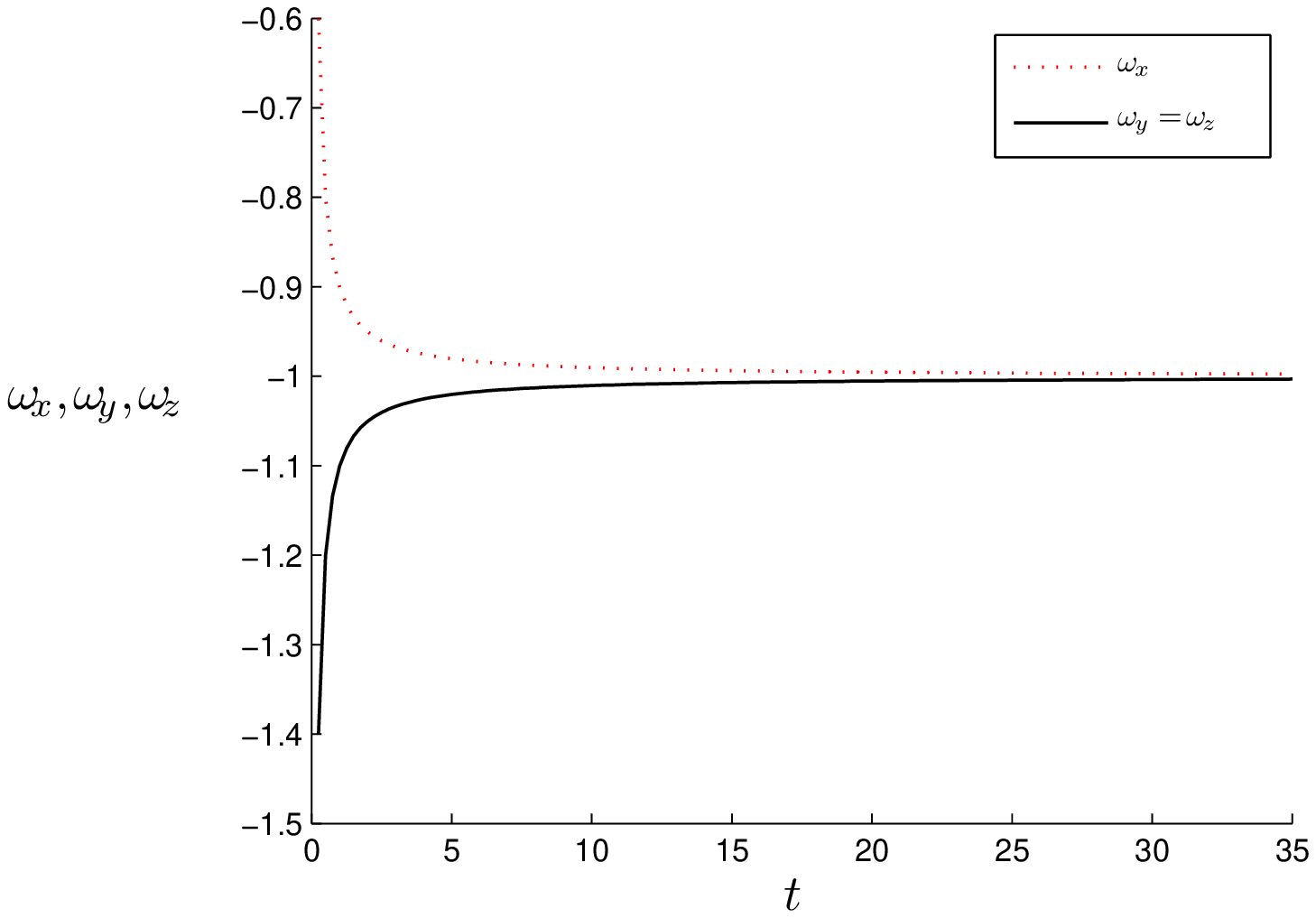}
\caption{Plot of directional EoS parameters versus time (t).} \label{fg:anil32F2.eps}
\end{center}
\end{figure}
\begin{figure}
\begin{center}
\includegraphics [height=6 cm]{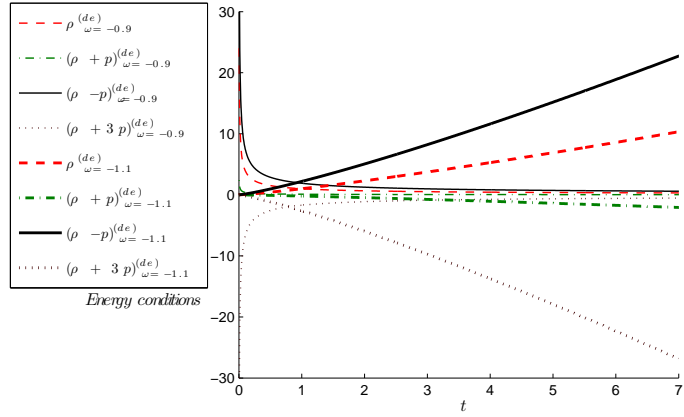}
\caption{Single plot of energy conditions} \label{fg:anil32F3.eps}
\end{center}
\end{figure}

From equations (\ref{eq14}) and (\ref{eq15}), the pressure and energy density of the perfect fluid are obtained as
\begin{eqnarray}\label{eq51}
p^{(m)}&=&(2n-3)(nt)^{-2}-l^{2}(nDt
)^{\frac{-6}{n}}-\frac{(2\omega+1)\alpha^{2}{\rho_{0}}^{2}}{3\omega^{2}}(nDt)^{\frac{-6(\omega+1)}{n}}\nonumber\\
& &
-\omega\rho_{0}(nDt)^{\frac{-3(\omega+1)}{n}}+\frac{\alpha^{2}}{(nDt)^{\frac{2}{n} } \exp \left[-\frac{4\alpha\rho_{0}}{3\omega
D(n-3\omega-3)}(nDt)^{\frac{n-3\omega-3}{n}} \right]} ,
\end{eqnarray}
\begin{eqnarray}\label{eq52}
\rho^{(m)}&=&3(nt)^{-2} -l^{2}(nDt
)^{\frac{-6}{n}}-\frac{\alpha^{2}{\rho_{0}}^{2}}{3\omega^{2}}(nDt)^{\frac{-6(\omega+1)}{n}} \nonumber\\
&
&
 -\frac{3\alpha^{2}}{(nDt)^{\frac{2}{n} } \exp \left[-\frac{4\alpha\rho_{0}}{3\omega
D(n-3\omega-3)}(nDt)^{\frac{n-3\omega-3}{n}} \right]}-\rho_{0}(nDt)^{\frac{-3(\omega+1)}{n}}.
\end{eqnarray}

\indent The perfect fluid density parameter $(\Omega^{(m)})$ and DE density parameter $(\Omega^{(de)})$ are given by
\begin{eqnarray}\label{eq53}
\Omega^{(m)}&=&1 -\frac{l^{2}(nDt
)^{\frac{-6}{n}}}{3(nt)^{-2}}-\frac{\alpha^{2}{\rho_{0}}^{2}}{9\omega^{2}(nt^{-2})}(nDt)^{\frac{-6(\omega+1)}{n}} \nonumber\\
&
&
 -\frac{1}{(nt)^{-2}}\left(\frac{3\alpha^{2}}{(nDt)^{\frac{2}{n} } \exp \left[-\frac{4\alpha\rho_{0}}{3\omega
D(n-3\omega-3)}(nDt)^{\frac{n-3\omega-3}{n}} \right]}+\rho_{0}(nDt)^{\frac{-3(\omega+1)}{n}}\right).
\end{eqnarray}

\begin{equation}
\label{eq54}
\Omega^{(de)} = \frac{\rho_{0}D^{-3(\omega+1)}}{n}\left(nt\right)^{\frac{2n-3(\omega+1)}{n}}
\end{equation}
Thus the overall density parameter $(\Omega)$ is obtained as
\begin{eqnarray}
\label{eq55}
\Omega&=&\Omega^{(m)}+\Omega^{(de)}\;\;\;\;\;\; \nonumber\\
&
&
= 1 -\frac{1}{3(nt)^{-2}}\left[l^{2}(nDt
)^{\frac{-6}{n}}-\frac{\alpha^{2}{\rho_{0}}^{2}}{3\omega^{2}}(nDt)^{\frac{-6(\omega+1)}{n}}
-\frac{3\alpha^{2}}{(nDt)^{\frac{2}{n} } \exp \left[-\frac{4\alpha\rho_{0}}{3\omega
D(n-3\omega-3)}(nDt)^{\frac{n-3\omega-3}{n}} \right]}\right].
\end{eqnarray}
\begin{figure}
\begin{center}
\includegraphics [height=6 cm]{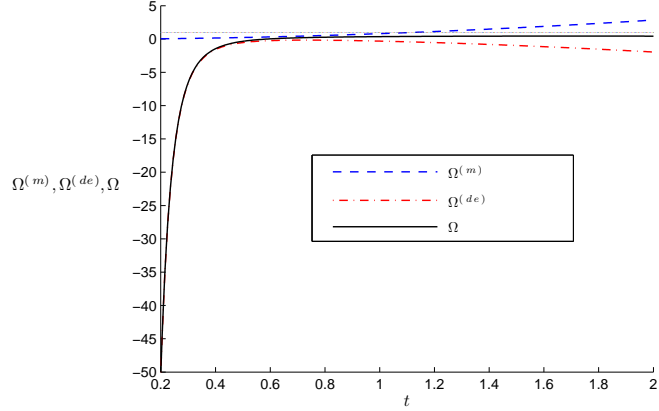}
\caption{Plot of density parameters versus time (t). } \label{fg:anil32RF1.eps}
\end{center}
\end{figure}
It is observed that at $t = 0$, the spatial volume vanishes while all other 
parameters diverge. Thus the derived model starts expanding with big bang singularity at $t = 0$. 
This singularity is point type because the directional scale factors $A(t)$, $(B(t))$ and $C(t)$ vanish 
at initial moment. The solutions for the scale factors have a combination of a
power-law term and exponential term in the product form. The DE term
appears in exponential form and thus affects their evolution
significantly. For $\alpha<0$ and $n>3\omega+3$, the DE contributes
to the expansion of $A(t)$ while opposing to the expansion of $B(t)$
and $C(t)$. Likewise, for $\alpha>0$  and $n>3\omega+3$, the
anisotropic DE opposes the expansion of $A(t)$ while contributing to
the expansion of $B(t)$ and $C(t)$.\\
\indent The difference between the directional EoS parameters and
hence the pressures of the DE, along x-axis and y-axis (or z-axis)
is $3\alpha(nt)^{-1}$, which decreases as $t$ increases. Therefore,
the anisotropy of the DE decreases as $t$ increases and finally
drops to zero at late time. The variation of mean anisotropic parameter $(\bar{A})$ versus 
has been graphed in $\bf{Fig. 1}$ by choosing $D = 2, \omega = -1.1, n = 0.5$ and other 
constant as unity. Since the current observations strongly recommend that the present 
Universe is accelerating (i. e. $ q < 0$ ). We consider $n = 0.5$ i. e. $q = -0.5$ in the 
remaining discussion of the model. \\
\indent $\bf{Fig. 2}$ depicts the variation of directional EoS parameter $(i. e. ~ \omega_{x}, 
~ \omega_{y}, ~ \omega_{z})$ 
versus cosmic time. We observe that directional EoS parameter along x-axis $(i. e. ~ \omega_{x})$ is 
decreasing function of time while directional EoS parameters along y-axis (or z-axis) are increasing 
function of time. At the later stage of evolution, all the directional EoS parameters approaches to $-1$ as expected. 
The same is predicted by current observations.\\

Following, Caldwell (2002), Srivastava (2005) and recently Yadav (2011) have investigated phantom model with 
$\omega < -1$. They have suggested that at late time, phantom energy has appeared as 
a potential DE candidate which violates the weak as well as strong energy condition. 
The left hand side of energy conditions have graphed in $\textbf{Fig. 3}$.\\
\indent From $\textbf{Fig. 3}$, 
for $\omega = -0.9$ (i. e. quintessence model), we observe that\\

(i) $\rho^{(de)}\geq 0$\\

(ii) $\rho^{(de)} + p^{(de)} \geq 0$\\

(iii) $\rho^{(de)} - p^{(de)} \geq 0$\\

(iv) $\rho^{(de)} + 3p^{(de)} < 0$\\

Thus the derived quintessence model violates the strong energy conditions, as expected.\\

Further, for $\omega = -1.1$ (i. e. phantom model), it is observed that\\

(i) $\rho^{(de)} > 0$\\

(ii) $\rho^{(de)} - p^{(de)} > 0$\\

(iii) $\rho^{(de)} + p^{(de)} < 0$\\

(iv) $\rho^{(de)} + 3p^{(de)} < 0$\\

Thus the derived phantom model violates the weak as well as strong energy conditions.
The same is predicted by current astronomical observations.\\

\indent From equation $(\ref{eq55})$, it is observed that for $0 < n < 1$ and $\omega > -1$,
the overall density parameter $(\Omega)$ approaches to 1 for sufficiently large 
times. Thus, the derived model predicts a flat Universe in quintessence model for sufficiently large times. 
{\bf Fig. 4}, depicts the variation of density
parameters versus cosmic time during the evolution of Universe.
\subsection{DE Cosmology for $n=0$}
In this case, integration of (\ref{eq16}) yields
\begin{equation}\label{eq56}
a(t)=c_{1}e^{Dt},
\end{equation}
where $c_{1}$ is a positive constant of integration.\\
The metric functions, therefore, read as
\begin{equation}\label{eq57}
A(t)=c_{1}\exp \left[Dt+\frac{2\alpha\rho_{0}} {9\omega
D(\omega+1)c_{1}^{3(\omega+1)}}e^{-3D(\omega+1)t} \right],
\end{equation}
\begin{equation}\label{eq58}
B(t)=m c_{1} \exp \left[Dt-\frac{l }{3Dc_{1}^{3} } e^{-3Dt} -\frac{\alpha\rho_{0}} {9\omega
D(\omega+1)c_{1}^{3(\omega+1)}}e^{-3D(\omega+1)t}\right],
\end{equation}
\begin{equation}\label{eq59}
C(t)=m^{-1} c_{1} \exp \left[Dt+\frac{l }{3Dc_{1}^{3} } e^{-3Dt}-\frac{\alpha\rho_{0}} {9\omega
D(\omega+1)c_{1}^{3(\omega+1)}}e^{-3D(\omega+1)t}
\right].
\end{equation}
provided $\omega\neq-1$. For $\omega=-1$, we have
\begin{equation}\label{eq60}
A(t)=c_{1}\exp \left[Dt+\frac{2\alpha\rho_{0}} {3}t \right],
\end{equation}
\begin{equation}\label{eq61}
B(t)=m c_{1} \exp \left[Dt-\frac{l }{3Dc_{1}^{3} } e^{-3Dt} -\frac{\alpha\rho_{0}} {3}t \right],
\end{equation}
\begin{equation}\label{eq62}
C(t)=m^{-1} c_{1} \exp \left[Dt+\frac{l }{3Dc_{1}^{3} } e^{-3Dt}-\frac{\alpha\rho_{0}} {3}t
\right].
\end{equation}
\indent The other cosmological parameters of the model have the
following expressions:
\begin{equation}\label{eq63}
H_{x}
=D-\frac{2\alpha\rho_{0}}
{3\omega c_{1}^{3(\omega+1)}}e^{-3D(\omega+1)t},
\end{equation}
\begin{equation}\label{eq64}
H_{y}=
D+\frac{\alpha\rho_{0}}
{3\omega c_{1}^{3(\omega+1)}}e^{-3D(\omega+1)t},
\end{equation}
\begin{equation}\label{eq65}
H_{z}
=D+\frac{\alpha\rho_{0}}
{3\omega c_{1}^{3(\omega+1)}}e^{-3D(\omega+1)t},
\end{equation}
\begin{equation}\label{eq66}
H=D,
\end{equation}
\begin{equation}\label{eq67}
\bar{A}= \frac{2\alpha^{2}{\rho_{0}}^{2}}{3 D^{2}\omega^{2}
c_{1}^{6(\omega+1)}} e^{-6D(\omega+1)t},
\end{equation}
\begin{equation}\label{eq68}
\theta =3D,
\end{equation}
\begin{equation}\label{eq69}
V =c_{1}^{3} e^{3Dt},
\end{equation}
\begin{equation}\label{eq70}
\sigma^{2} = \frac{\alpha^{2}{\rho_{0}}^{2}}{3\omega^{2}
c_{1}^{6(\omega+1)}} e^{-6D(\omega+1)t}.
\end{equation}
\indent The skewness parameters and the directional EoS parameters
of DE, respectively, are given by
\begin{equation}\label{eq71}
\delta(t)=\frac{\alpha c_{1}^{3(\omega+1)}}{\rho_{0}}\left[2D e^{3D(\omega +1)t}+\frac{2\alpha\rho_{0}}
{3\omega c_{1}^{3(\omega+1)}}\right],
\end{equation}
\begin{equation}\label{eq72}
\gamma(t)=\eta(t)=-\frac{\alpha c_{1}^{3(\omega+1)}}{\rho_{0}}\left[D e^{3D(\omega +1)t}-\frac{2\alpha\rho_{0}}
{3\omega c_{1}^{3(\omega+1)}}\right],
\end{equation}
\begin{equation}\label{eq73}
\omega_{x}=\omega+\frac{\alpha c_{1}^{3(\omega+1)}}{\rho_{0}}\left[2D e^{3D(\omega +1)t}+\frac{2\alpha\rho_{0}}
{3\omega c_{1}^{3(\omega+1)}}\right],
\end{equation}
\begin{equation}\label{eq74}
\omega_{y}=\omega_{z}=\omega-\frac{\alpha c_{1}^{3(\omega+1)}}{\rho_{0}}\left[D e^{3D(\omega +1)t}-\frac{2\alpha\rho_{0}}
{3\omega c_{1}^{3(\omega+1)}}\right],
\end{equation}
\begin{figure}
\begin{center}
\includegraphics [height=6 cm]{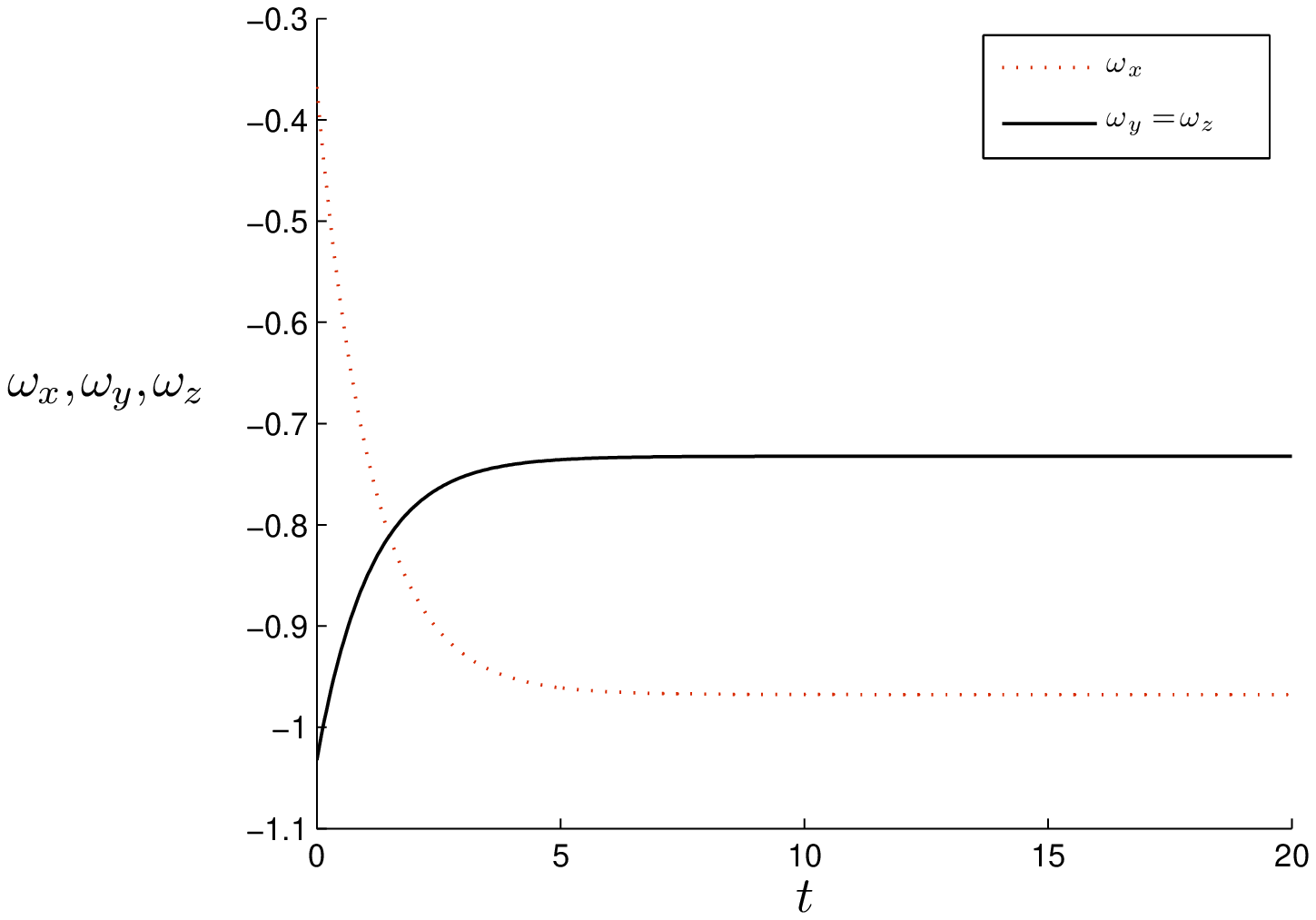}
\caption{Plot of directional EoS parameters versus time (t).} \label{fg:anil32F4.eps}
\end{center}
\end{figure}
\begin{figure}
\begin{center}
\includegraphics [height=6 cm]{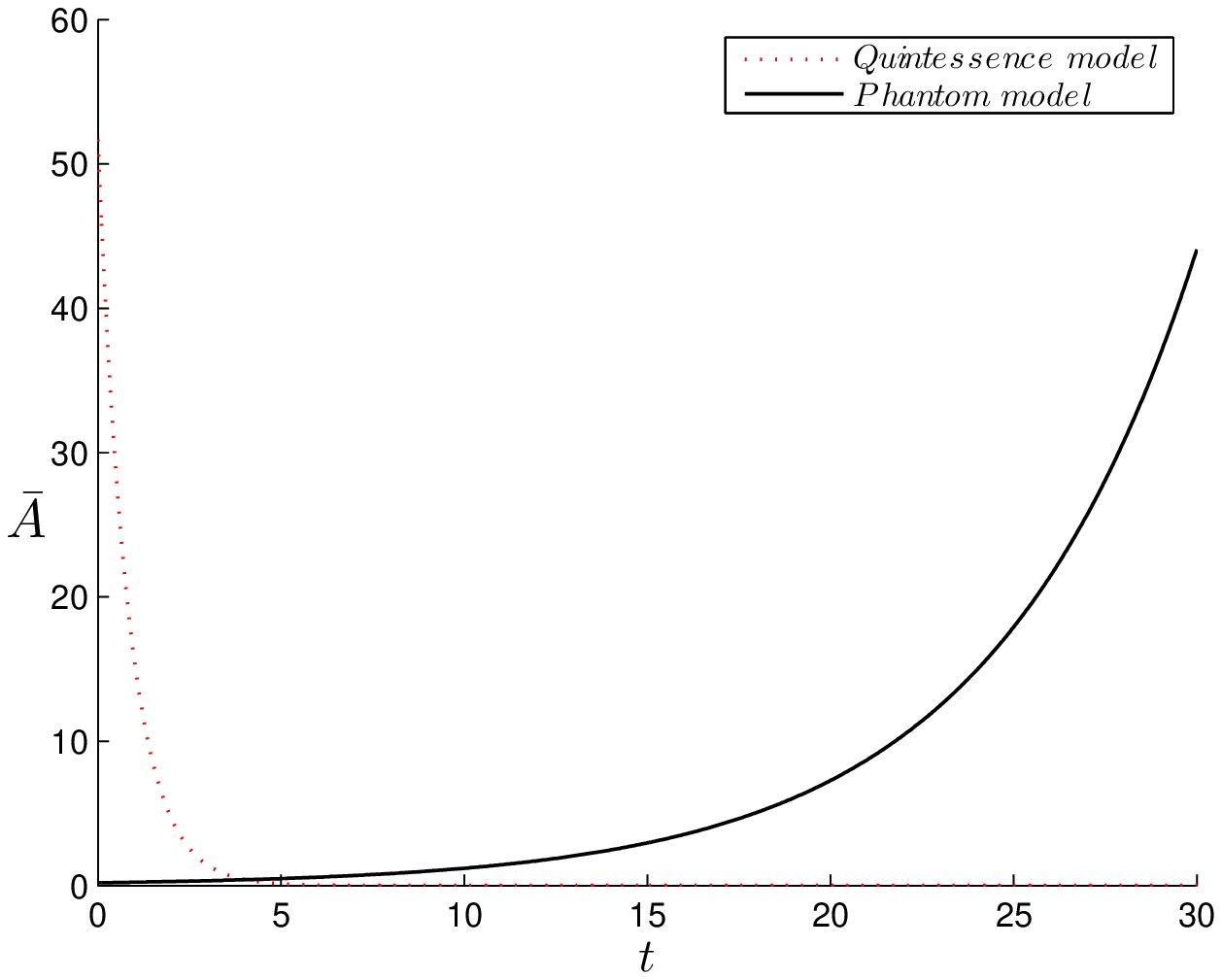}
\caption{Plot of Anisotropic parameter $(\bar{A})$ versus time (t). for n=0} \label{fg:anil32F5.eps}
\end{center}
\end{figure}
\begin{figure}
\begin{center}
\includegraphics [height=6 cm]{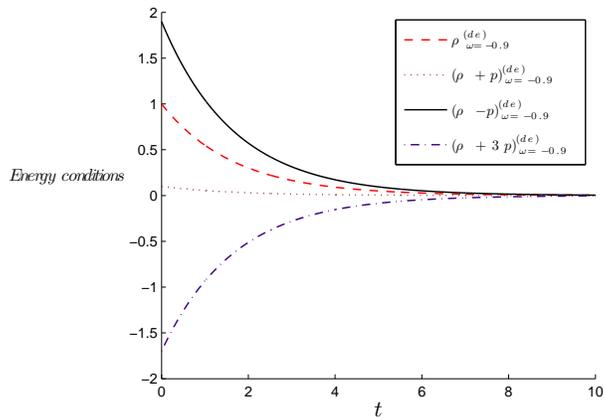}
\caption{Single plot of energy conditions in quintessence model for n = 0} \label{fg:anil32F6.eps}
\end{center}
\end{figure}
\begin{figure}
\begin{center}
\includegraphics [height=6 cm]{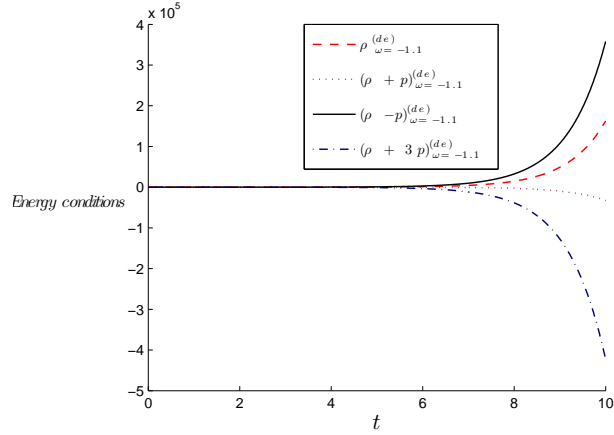}
\caption{Single plot of energy conditions in phantom model for n = 0.} \label{fg:anil32F7.eps}
\end{center}
\end{figure}

\indent The energy density and pressure of the DE components are
given by
\begin{equation}\label{eq75}
\rho^{(de)}=\rho_{0}c_{1}^{-3(\omega+1)}e^{-3D(\omega+1)t},
\end{equation}
\begin{equation}\label{eq76}
p^{(de)}=\omega\rho_{0}c_{1}^{-3(\omega+1)}e^{-3D(\omega+1)t}.
\end{equation}
From equations (\ref{eq14}) and (\ref{eq15}), the pressure and energy density of the perfect fluid components are obtained as
\begin{equation}\label{eq77}
p^{(m)}=-3D^{2}-\frac{l^{2}}{c_{1}^{6}}
e^{-6Dt}-\frac{(2\omega+1)
\alpha^{2}{\rho_{0}}^{2}}{3\omega^{2}c_{1}^{6(\omega+1)}}e^{-6D(\omega+1)t}
-\frac{\omega\rho_{0}}{c_{1}^{3(\omega+1)}}e^{-3D(\omega+1)t},
\end{equation}
\begin{equation}\label{78}
\rho^{(m)}=3D^{2}-\frac{l^{2}}{c_{1}^{6}}
e^{-6Dt}-\frac{\alpha^{2}{\rho_{0}}^{2}}{3\omega^{2}c_{1}^{6(\omega+1)}}e^{-6D(\omega+1)t}
-\frac{\rho_{0}}{c_{1}^{3(\omega+1)}}e^{-3D(\omega+1)t}.
\end{equation}
\indent The perfect fluid density parameter $(\Omega^{(m)})$ and DE density parameter $(\Omega^{(de)})$ are given by
\begin{equation}
\label{eq79}
\Omega^{(m)}=1-\frac{1}{3D^2}\left[\frac{l^{2}}{c_{1}^{6}}
e^{-6Dt}+\frac{\alpha^{2}{\rho_{0}}^{2}}{3\omega^{2}c_{1}^{6(\omega+1)}}e^{-6D(\omega+1)t}
+\frac{\rho_{0}}{c_{1}^{3(\omega+1)}}e^{-3D(\omega+1)t}\right]
\end{equation}
\begin{equation}
\label{eq80}
\Omega^{(de)} = \frac{\rho_{0}c_{1}^{-3(\omega+1)}}{3D^{2}} e^{-3D(\omega+1)t}
\end{equation}
Thus the overall density parameter $(\Omega)$ is obtained as
\begin{equation}
\label{eq81}
\Omega = 1-\frac{1}{3D^2}\left[\frac{l^{2}}{c_{1}^{6}}
e^{-6Dt}+\frac{\alpha^{2}{\rho_{0}}^{2}}{3\omega^{2}c_{1}^{6(\omega+1)}}e^{-6D(\omega+1)t}\right]
\end{equation}

We observe that the model has no initial singularity. The
directional scale factors and all other physical quantities are
constants at $t=0$. The directional scale factors and volume of the
universe increase exponentially with the cosmic time whereas the
mean Hubble parameter and expansion scalar are constants throughout
the evolution. Therefore, uniform exponential expansion takes place.
Further, we see that the DE term appears in exponential form in the
scale factors and thus effects their evolution significantly.
Thus, the spatial geometry of the universe is affected by the
anisotropic DE. The difference between the directional EoS
parameters of the DE and hence the pressures of the DE, along x-axis
and y-axis (or z-axis) is $3\alpha D$, which is constant throughout
the evolution of the universe. Therefore, the anisotropy of the DE does not 
vanish during the evolution of universe.\\
\indent As  $\;t\to \infty $, the scale factors and volume of the      
universe become infinitely large whereas the skewness and
directional EoS parameters, directional Hubble parameters become constants. The pressure
and energy density of the DE drops to zero in quintessence model (i. e. $\omega > -1$). 
$\textbf{Fig. 6}$ shows the plot of anisotropy parameter ($\bar{A}$) versus time, in both 
quintessence model and phantom model. We see that in quintessence model, the anisotropy parameter 
decreases as time increases and finally drops to zero at late time. But in phantom model 
the anisotropy parameter does not vanish during the evolution of Universe. 
The left hand side of energy conditions have graphed in $\textbf{Fig. 7 and Fig. 8}$ for quintessence model 
and phantom model respectively. It is observed that the quintessence model violates strong energy condition 
whereas phantom model violate weak energy condition as well as strong energy condition, as expected.\\

\indent For $\;n=0$, we get $\;q=-1$; incidentally this value of
deceleration parameter leads to $\;\frac{dH}{dt}=0$, which implies
the greatest value of Hubble's parameter and the fastest rate of
expansion of the universe. Therefore, the derived model can be utilized 
to describe the dynamics of the late time evolution of the actual Universe. So, 
in what follows, we emphasize upon the late time behaviour of the derived model. 
$\bf{Fig. 5}$ depicts the variation of directional EoS parameter versus time for $n = 0$. 
We observe that $\omega_{x}, \omega_{y}$ or $\omega_{z}$ are evolving with negative sign, as expected.\\ 
\indent From equation (\ref{eq81}), it is observed that for 
sufficiently large time, the overall density parameter $(\Omega)$ approaches to $1$ 
in quintessence model. Thus the derived model predicts a flat Universe at late time.\\
\section{Concluding Remarks}
In this paper, some spatially homogeneous and anisotropic DE models in 
Bianchi type V space-time have been studied. The main features of the work are as follows

\begin{itemize}

\item The models are based on the exact solution of Einstein's field equations for 
anisotropic Bianchi type V space-time filled with DE.

\item The singular model $(n\neq 0)$ seems to describe the dynamics of the Universe 
from big bang to the present epoch while non singular model $(n = 0)$ seems reasonable 
to project dynamics of the future Universe.

\item The directional EoS parameters (i. e. $\omega_{x}, ~ \omega_{y} \; or \; \omega_{z}$) evolve with in the range 
predicted by observations.

\item In the present models, we do not rule out the anisotropic nature of DE. 
The anisotropic DE contributes to the expansion
of one (or two) of the scale factors while it opposes the expansion
of the other two (or one) scale factors leading the geometry of Universe. 
The anisotropy of DE vanishes at late time for singular model (see $\textbf{Fig. 1}$) whereas for 
non singular model, anisotropy occurs at early stage (i. e. quintessence model) or at later time of Universe 
(i. e. phantom model) (see $\textbf{Fig. 6}$).

\item The derived quintessence models violate the strong energy condition whereas the phantom models 
violate the weak energy condition as well as strong energy condition (see $\textbf{Fig. 3, Fig. 7, Fig. 8}$).

\item The flatness of Universe can be achieved in quintessence model $(i. e.\;\; \omega > -1)$ for sufficiently large time 
because the overall density parameter $(\Omega)$ approaches to $1$. Thus in our analysis, the quintessence model is 
turning out as a suitable model for describing the late time evolution of Universe.
 
\end{itemize}

\section*{Acknowledgements} 
Author would like to thanks the anonymous learned referee for his/her valuable comments which 
improved the paper in this form. Also Author thanks to S. Kumar for helpful 
discussions.


\end{document}